\documentstyle[fleqn,epsf,graphicx]{article}
\begin{document}
\title{The BCS-BEC crossover and superconductivity in a lattice fermion model with hard core repulsion}
\author{Agnieszka Kujawa}
\date{}

\maketitle

\vspace{-1.1cm}
\begin{center}
\begin{footnotesize}
\emph{Solid State Theory Division, Faculty of Physics,\\ Adam Mickiewicz University,\\ Umultowska 85,
61-614 Pozna\'n, Poland}
\end{footnotesize}
\end{center}

\begin{center}
\begin{footnotesize}
\end{footnotesize}
\end{center}

\begin{abstract}
We have considered the BCS-BEC crossover at $T=0$ in the framework of a lattice fermion model (the extended Hubbard model). We have analyzed the case of on-site hard core repulsion ($U=\infty$) and intersite attraction ($W<0$) in a 3D system with extended $s$-wave pairing symmetry. The mechanisms of density-driven and attraction-strength-driven crossover have been analyzed. We have obtained the ground state phase diagram. The analyzed quantities are: chemical potential, order parameter, coherence length and fraction of condensed particles. 
\end{abstract}

\section{Introduction}
High-$T_c$ superconductivity has been widely discussed since its discovery by Bednorz and M\"uller in 1986 \cite{bednorz}. A microscopic mechanism which could explain the characteristic properties of High-$T_c$ superconductors has not been found yet. However, there has been broad agreement that the physics of the BCS-BEC crossover is crucial for High-$T_c$ superconductivity \cite{MicnasModern}. The effects of the crossover are clearly seen in the behaviour of the energy gap and the chemical potential. The so-called pseudogap has been observed in the excitation spectra of many unconventional superconductors \cite{chen}. 

In 2004 Jin's group experimentally observed a condensation of pairs of fermionic atoms in the region of the BCS-BEC crossover \cite{jin}. A trapped gas of fermionic atoms of $^{40}$K was cooled to suitably low temperatures of the order of $5\cdot 10^{-8}$ K and thereafter the interatomic interaction strength was controlled via the Feshbach resonance.

In this paper we analyze the physics of the BCS-BEC crossover in High-$T_c$ superconductors by means of a lattice fermion (the extended Hubbard) model for the (3D) simple cubic lattice system with hard core on-site repulsion and intersite attraction. This paper consists of three parts. The first part gives a discussion of the extended Hubbard model in the Hartree-Fock approximation. The second part presents the properties of selected quantities and the numerical results. In sec. III we summarize the discussion. 

\section{Model}
The model Hamiltonian is the extended Hubbard model with on-site ($U$) and intersite ($W$) interactions:
\cite{nozieres}, \cite{Micnas10lat}, \cite{Micnas1988}:
\begin{equation}
\label{extham}
H=\sum_{ij}\sum_{\sigma}t_{ij}c_{i\sigma}^{\dag}c_{j\sigma}+\frac{1}{2}U\sum_{i\sigma}n_{i\sigma}n_{i-\sigma}+
\frac{1}{2}\sum_{ij}\sum_{\sigma \sigma'}W_{ij}n_{i\sigma}n_{j\sigma'},
\end{equation}
where $n_{i\sigma}=c_{i\sigma}^{\dag}c_{i\sigma}$, $t_{ij}$ -- hopping integral. Having transformed Hamiltonian (\ref{extham}) to the reciprocal lattice and using the Hartree-Fock approximation, the following system of equations for the superconducting order parameter ($\Delta_k=\Delta_0+\gamma_k \Delta_{\gamma}$) in the case of the $s^*$ pairing symmetry is obtained:
\begin{equation}
\label{sext}
\left(\begin{array}{ccc}
1+U\phi_1(T) & U\phi_2(T) \\
-\frac{\vert W \vert}{z}\phi_2(T) & 1-\frac{\vert W \vert}{z}\phi_{\gamma}(T) \\
\end{array}\right)
\left(\begin{array}{ccc}
\Delta_0 \\
\Delta_{\gamma} \\
\end{array} \right) =0,
\end{equation}
where: $\phi_1(T)=\frac{1}{N}\sum_qF_q(T)$, $\phi_2(T)=\frac{1}{N}\sum_q\gamma_q F_q(T)$, $\phi_{\gamma}(T)=\frac{1}{N}\sum_q\gamma_q^2 F_q(T)$, $F_q=(2E_q)^{-1}\textrm{tanh} (\beta E_q \slash 2)$, $\gamma_k=2(\textrm{cos}k_xa_x + \textrm{cos}k_ya_y +\textrm{cos}k_za_z )$, $a_i$ -- lattice constant in the $i$-th direction (we set $a_x=a_y=a_z=1$ in further considerations), $z$ -- coordination number.
The equation for the chemical potential takes the form:
\begin{equation}
\label{pch}
n-1=-\frac{2}{N}\sum_k\bar{\epsilon_k}F_k(T),\vspace{-0.5cm}
\end{equation}
where $\bar{\epsilon_k}=\epsilon_k-\bar{\mu}$, $n$ -- electron concentration.
Solving the above equations simultaneously, one can find the dependence of the energy gap and the chemical potential on the electron concentration. Taking into account only the hopping between the nearest neighbours, the electron dispersion is: 
\begin{equation}
\epsilon_k=-2t(\textrm{cos}k_x+\textrm{cos}k_y+ \textrm{cos}k_z).
\end{equation}
The evolution from the weak coupling limit (BCS) to the limit of the strong coupling (BEC) takes place when we decrease the electron concentration or increase the interaction. According to the Leggett criterion \cite{Leggett}, the BCS-BEC crossover takes place when the modified chemical potential: 
\begin{equation} 
\bar{\mu}=\mu-n(U\slash 2+W_0)
\end{equation}
drops below the lower band edge.

When $U\rightarrow \infty$, the system of equations for the order parameter takes the form (provided that $\Delta_0 \neq 0$, $\Delta_{\gamma} \neq 0$):
\begin{equation}
\label{Uinf}
\Phi_1(T) \Delta_0+\Phi_2 (T) \Delta_{\gamma}=0,    
\end{equation}
\begin{equation}
-\frac{\vert W \vert}{z} \Phi_2(T) \Delta_0 + \left( 1-\frac{\vert W \vert}{z} \Phi_{\gamma}(T) \right) \Delta_{\gamma}=0. 
\end{equation}
Because of the fact that we consider the case of $U\slash t=\infty$, the on-site pairing is forbidden. One can show it by calculating the anomalous corelation function \cite{Pistolesi'}.

\section{Results}

\begin{figure}
\begin{center}
\includegraphics[width=0.34\textwidth,angle=270]{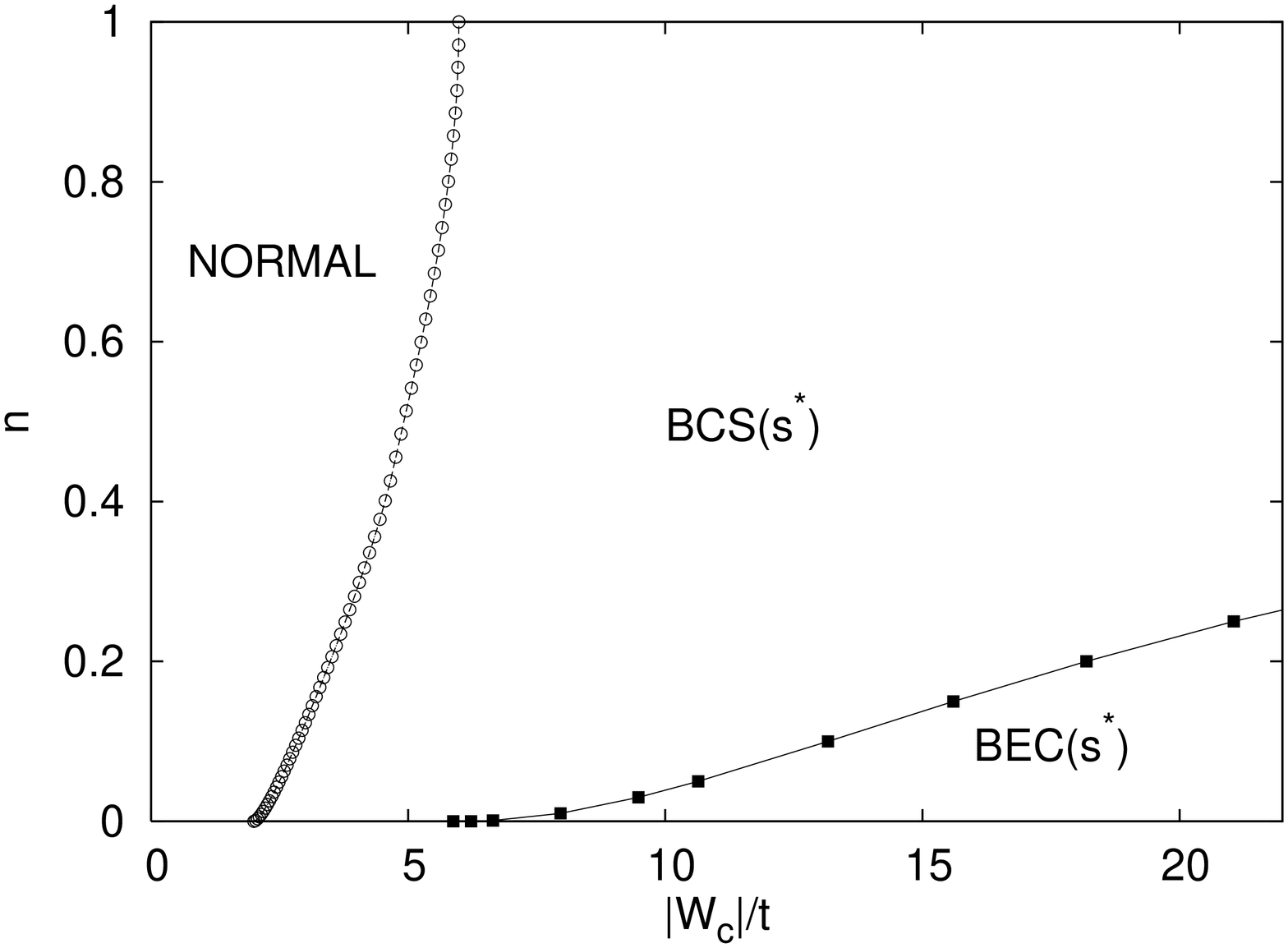}
\includegraphics[width=0.34\textwidth,angle=270]{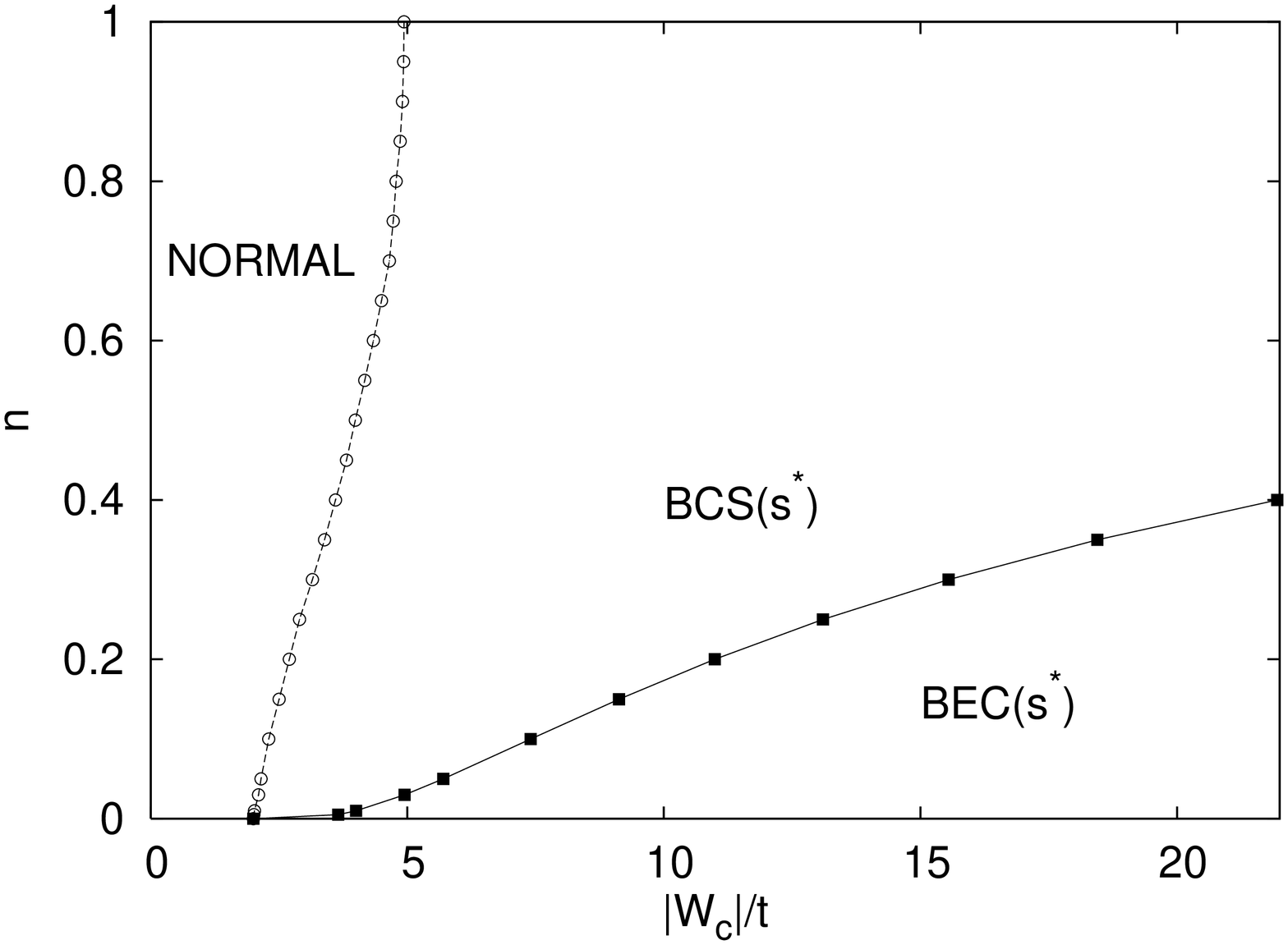}
\caption{The ground state phase diagrams ($U=\infty$) for the 3D case (left) and the 2D case (right).}
\end{center}
\end{figure}

We have performed an analysis of the BCS-BEC crossover for the model with hard core repulsion. The analyzed quantities are: chemical potential, order parameter, coherence length and fraction of condensed particles.

Fig. 1 shows the ground state ($T=0$) phase diagram for the 3D case (left) and the 2D case (right; this case has been analyzed in detail in Ref \cite{Pistolesi'}). Both in the 2D and the 3D case there exists a critical interaction value for which a bound state forms in the empty lattice (when $n\rightarrow 0$). For the 2D case $|W_c|\slash t=2$. At the same time, it is the minimal interaction value for which a transition to the normal state takes place.
The very existence of the transition to the normal state at $T=0$ at suitably low values of the intersite interaction is interesting -- it indicates that quantum fluctuations are very strong.
For the 3D case the critical interaction value for which a bound state forms in the $n\rightarrow 0$ limit is $|W_c|\slash t\approx5.874$ but the critical interaction value for which a transition to the normal state takes place is $|W_c|\slash t=2$. Therefore, as oposed to the 2D case, the minimal interaction value for the emergence of superconductivity is different than the critical interaction value for the emergence of a bound state in the empty lattice.
In both cases the range of occurrence of the local pairs phase widens when we decrease the electron concentration or we increase the interaction. Because of the fact that we consider the case of $U\slash t= \infty$, the maximum value of the electron concentration is $n=1$.

Now, let us analyze the BCS-BEC crossover in the context of the behaviour of the order parameter and the chemical potential. The left panel of Fig. 2 shows the dependence of the order parameter at the Fermi level (then, \mbox{$\Delta_k=\Delta_0-(\bar{\mu}\slash t)\Delta_{\gamma})$} on the electron concentration for three values of the intersite interaction. For low electron concentrations this is a square root dependence, characteristic of the BEC limit. However, for higher electron concentrations one can observe the exponential drop of
$\Delta_k(\bar{\mu})\slash t$, characteristic of the BCS limit.

\begin{figure}
\begin{center}
\includegraphics[width=0.34\textwidth,angle=270]{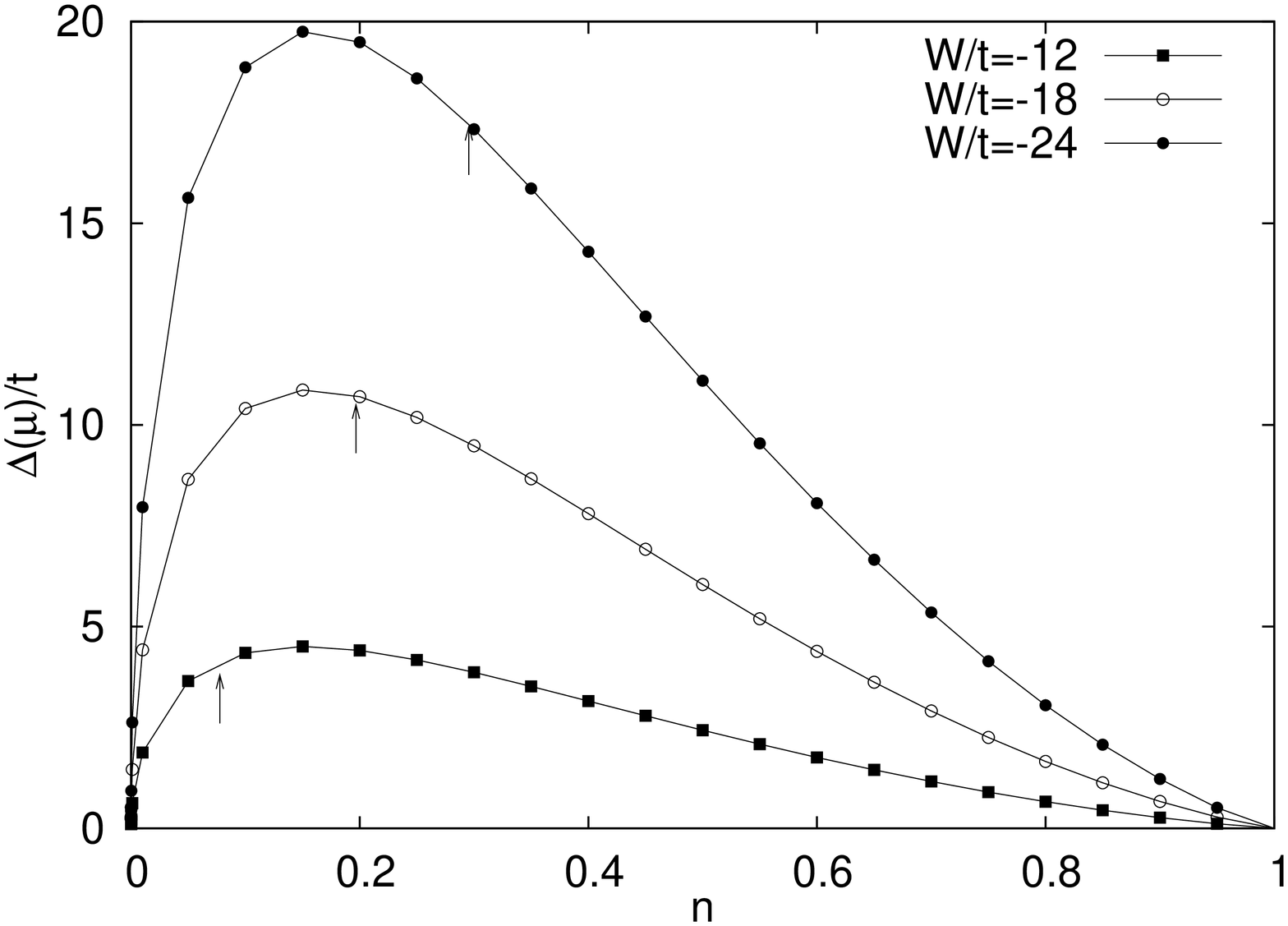}
\includegraphics[width=0.34\textwidth,angle=270]{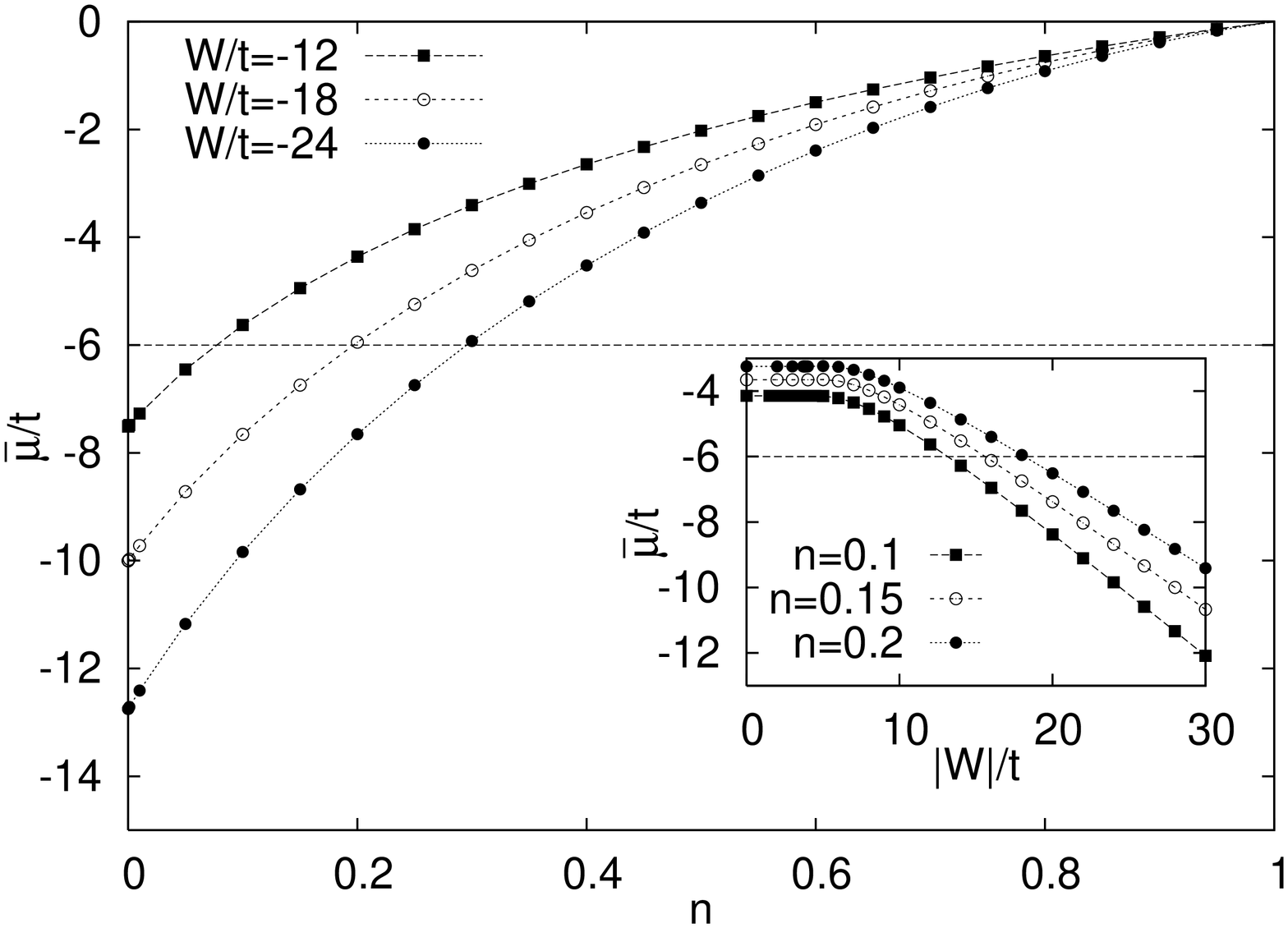}
\caption{(left) Order parameter at the Fermi level vs. electron concentration; the arrows show the BCS-BEC crossover points. (right) The dependence of the modified chemical potential on the electron concentration and the intersite interaction $|W|$ (inset); the horizontal dashed line denotes the band bottom.}
\end{center}
\end{figure}

\begin{figure}
\begin{center}
\includegraphics[width=0.5\textwidth,angle=270]{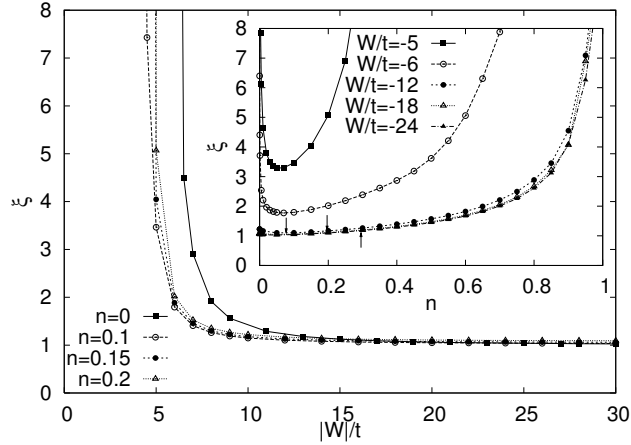}
\caption{The dependence of the coherence length in units of the lattice constant on the intersite interaction and the electron concentration (inset); the arrows show the BCS-BEC crossover points.}
\end{center}
\end{figure}

The right panel of Fig. 2 shows the dependence of the modified chemical potential on the electron concentration and the intersite interaction (inset). As we mentioned before, according to the Leggett criterion \cite{Leggett}, the BCS-BEC crossover takes place when the modified chemical potential drops below the lower band edge (in the 3D case the limiting value is $\bar{\mu}/t=-6$). The horizontal dashed line denotes the band bottom.
The values of $\bar{\mu}$ for $n=0$ are exact  (despite the fact that we have used the Hartree-Fock approximation) and correspond to one half of the bound state energy. 

A very important quantity from the point of view of the BCS-BEC crossover is the coherence length $\xi$. The suggestions to study this quantity in this context result from the observations of the properties of exotic superconductors. In agreement with the Uemura's plot \cite{Uemura}, the unconventional superconductors have a considerably shorter coherence length $\xi$ ($\sim 20-50$ \AA $\,$) than the conventional superconductors ($\xi \sim 10^3-10^4$ \AA $\,$). The coherence length (pair size) can be defined as:
\begin{equation}
\xi^2=\frac{\sum_k \vert\vec{\nabla_k} \phi_k \vert^2}{\sum_k \vert \phi_k \vert^2},
\end{equation}
where $\phi_k=u_k\nu_k=\Delta_k \slash 2 E_k$ is the pair wave function in the ground state.

Fig. 3 shows the dependence of the coherence length in units of the lattice constant on the intersite interaction and the electron concentration (inset).
Because of the fact that we consider the case of hard-core repulsion and intersite attraction, $\xi$ exceeds the value of the lattice constant for any interaction values.
For the values of $|W|/t$ which slightly exceed the critical value for a given electron concentration (5.874 for $n=0$), the pair size for small $n$ can be as big as the pair size in the BCS regime.
When we add electrons, the Pauli Exclusion Principle causes a decrease of the pair size, which then becomes just over 1.
Along with a further increase of the electron concentration (when a crossover to the BCS regime for a given interaction strength occurs), the pairs can have large sizes (due to the pairing in the momentum space).
Fig. 3 shows also the dependence of the coherence length on the intersite interaction. For each value of $n$, the asymptote of the function $\xi$ is a line coresponding to the critical interaction value for which a transition to the normal state takes place.

Another quantity which describes the BCS-BEC crossover quantitatively is the fraction of condensed particles:
\begin{equation}
\frac{n_0}{n}=\frac{\frac{1}{N} \sum_k \left( \frac{\Delta_k}{2E_k} \right)^2}{n}.
\end{equation}

\begin{figure}
\begin{center}
\includegraphics[width=0.5\textwidth,angle=270]{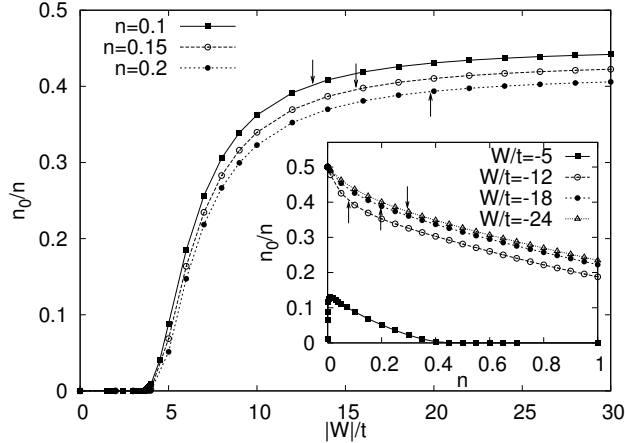}
\caption{The dependence of the fraction of condensed particles on the intersite interaction and electron concentration (inset); the arrows show the BCS-BEC crossover points.}
\end{center}
\end{figure}

Fig. 4 shows the dependence of the fraction of condensed particles on the intersite interaction and electron concentration (inset). When we increase the intersite interaction or decrease the electron concentration, the fraction of condensed particles increases. In the empty lattice limit, $n_0/n$ takes the largest value. The maximum value of the ratio $n_0/n$ is 0.5, because if all particles in the system form pairs and condensate, their concentration is twice lower than the initial electron concentration. A very interesting example is the case of $|W|/t=5$. Because of the fact that this interaction value does not exceed the critical interaction value for pair binding, the saturation of the fermion pairs in the condensate is very small, as the system is in the BCS regime.

The case of the fraction of condensed particles exemplifies very well that the mean-field approximation fails in the regime of higher electron concentration in the hard core repulsion case. Because of the fact that $U/t=\infty$, in the $n\rightarrow 1$ limit one should observe a transition to the Mott insulator phase. Indeed, Fig. 4 shows that the fraction of condensed particles is too high in the $n\rightarrow 1$ limit.

\section{Conclusions}
The mechanisms of density-driven and attraction-strength-driven BCS-BEC crossover have been considered. The range of occurrence of the local pairs phase widens when we decrease the electron concentration $n$ or we increase the attractive interaction $|W|$. We have considered the $s^*$-wave pairing, for which the evolution from the weak coupling limit (BCS) to the limit of the strong coupling (BEC) is smooth. For the case of $U=\infty$, both in the 2D and the 3D case there exists a critical interaction value for which a bound state forms in the empty lattice (when $n\rightarrow 0$) and the critical interaction value for which a transition to the normal state takes place. The two critical values coincide in the 2D case, but in the 3D case the critical value for the emergence a bound state is higher than for the emergence of superconductivity. We have noticed that despite the fact that we have used the Hartree-Fock approximation, the values of $\bar{\mu}$ for $n=0$ are exact and correspond to one half of the energy of the bound state. We have shown that the pair size for small $n$ can be as large as the pair size in the BCS regime. Indeed, the Hartree-Fock approximation fails for higher electron concentrations in the hard core repulsion case -- we have considered this issue in the context of the behaviour of the fraction of condensed particles.

\section*{Acknowledgements}
The results of this paper are a part of the M.Sc. thesis, completed under the supervision of Prof. Roman Micnas, to whom I am very grateful for guidance and valuable discussions.
A partial support from the Polish Science Foundation is acknowledged.

\end{document}